\documentclass{aastex}
\usepackage{emulateapj5}
\usepackage{amssymb}
\makeatletter

\renewcommand{\cite}[1]{\citeyear{#1}}
\bibpunct{\hspace{-4pt}}{}{}{a}{}{}
\submitted{{\rm Accepted to ApJ Lett.}}
\newcounter{twocoltable}[table]
\newcommand{\twocoltablecap}[2]{%
   \refstepcounter{twocoltable}%
   \label{#1}%
   \begin{center}%
   {\sc \small%
   Table \thetwocoltable \\%
   \vspace{1ex}
   #2}%
   \end{center}%
}
\makeatother
\begin{document}

\newcommand{\Ua}{\widetilde{a}}

\newcommand{\UP}{\widetilde{P}}

\newcommand{\Um}{\widetilde{m}}

\newcommand{\Ut}{\widetilde{t}}

\newcommand{\Urp}{\widetilde{r}_{p}}

\newcommand{\UdE}{\Delta\widetilde{E}}
 
\newcommand{\Urt}{\widetilde{r}_{t}}

\newcommand{\Urh}{\widetilde{r}_{h}}

\newcommand{\UE}{\widetilde{E}}

\newcommand{\UR}{\widetilde{R}}

\newcommand{\Mo}{M_{\odot}}

\newcommand{\Ro}{R_{\odot}}
 
\newcommand{\Lo}{L_{\odot}}

\newcommand{\HLm}{\widehat{L}_{0}}

\newcommand{\Utm}{\widetilde{t}_{0}}
 
\newcommand{\UT}{\widehat{T}}

\newcommand{\ULt}{\widetilde{L}_{t}}

\newcommand{\HLt}{\widehat{L}_{t}}

\newcommand{\Utd}{\widetilde{\tau}_{d}}

\newcommand{\Ms}{M_{\star}}

\newcommand{\Rs}{R_{\star}}
 
\newcommand{\Ls}{L_{\star}}

\newcommand{\Ts}{T_{\star}}

\newcommand{\Ns}{N_{\star}}
 
\newcommand{\UWp}{\widetilde{\Omega}_{p}}
 
\newcommand{\UdJ}{\Delta\widetilde{J}}

\newcommand{\Te}{T_{\mathrm{eff}}}

\newcommand{\Uc}{\widetilde{c}}

\title{Orbital inspiral into a massive black hole in a galactic center}

\author{Tal Alexander and Clovis Hopman }

\affil{Faculty of Physics, The Weizmann Institute of Science, POB 26, Rehovot
76100, Israel}

\begin{abstract}
A massive black hole (MBH) in a galactic center drives a flow of stars
into nearly radial orbits to replace those it destroyed. Stars whose
orbits cross the event horizon $r_{S}$ or the tidal disruption radius
$r_{t}$ are promptly destroyed in an orbital period $P$. Stars with
orbital periapse $r_{p}$ slightly larger than the sink radius $q\!\equiv\!\max(r_{S},r_{t})$
may slowly spiral in due to dissipative interactions with the MBH,
e.g. gravitational wave emission, tidal heating or accretion disk
drag, with observable consequences and implications for the MBH growth
rate. Unlike prompt destruction, the inspiral time is typically $\gg\! P$.
This time is limited by the same scattering process that initially
deflected the star into its eccentric orbit, since it can deflect
it again to a wider orbit where dissipation is inefficient. The ratio
between slow and prompt event rates is \emph{}therefore \emph{much
smaller} than that implied by the ratio of cross-sections, $\sim\! r_{p}/q$,
and so only prompt disruption contributes significantly to the mass
of the MBH. Conversely, most stars that scatter off the MBH survive
the extreme tidal interaction ({}``tidal scattering''). We derive
general expressions for the inspiral event rate and the mean number
of inspiraling stars, and show that the survival probability of tidally
scattered stars is $\sim\!1$, and that the number of tidally heated
stars ({}``squeezars'') and gravity wave emitting stars in the Galactic
Center is $\sim\!0.1$--$1$. 
\end{abstract}


\section{Introduction}

\label{sec:Introduction}

Evidence for the presence of massive black holes in the centers of
most galaxies (e.g. Gebhardt et al. \cite{Geb03}), together with
recent and anticipated advances in observing capabilities, have focused
interest on the observational implications of star-MBH interactions
such as tidal disruption (Frank \& Rees \cite{Fra76}; Frank \cite{Fra78};
Lightman \& Shapiro \cite{Lig77}; Syer \& Ulmer \cite{Sye99}; Magorrian
\& Tremaine \cite{Mag99}), tidal scattering (Alexander \& Livio \cite{Ale01b}),
gravitational wave (GW) emission (e.g. Hils \& Bender \cite{Hil95};
Sigurdsson \& Rees \cite{Sig97a}; Freitag \cite{Fre01}, \cite{Fre03}),
interaction with a massive accretion disk surrounding the MBH (Ostriker
\cite{Ost83}; Syer, Clarke \& Rees \cite{Sye91}; Vilkoviskij \&
Czerny \cite{Vil02}), and tidal capture and tidal heating (Novikov,
Petchik \& Polnarev \cite{Nov92}; Alexander \& Morris \cite{Ale03},
AM03). Much effort has been devoted to the study of tidal disruption,
which plays an important role in the growth of low-mass MBHs (Murphy,
Cohn \& Durisen \cite{Mur91}; Freitag \& Benz \cite{Fre02a}) and
can provide a signature for the existence of MBH in galactic nuclei
by the emission of tidal flares.

Dynamical analyses indicate that most of the stars scattered into
radial orbits originate at the MBH radius of influence, $r_{h}$,
where the enclosed stellar mass roughly equals the MBH mass $m$ and
the scattering is roughly isotropic. Event horizon crossing or tidal
disruption is prompt; the stars plunge toward the MBH with a cross-section
that scales as $\sim\! r_{p}$ (Hills \cite{Hil75}) and reach it
in less than the initial orbital period, $P_{0}$, irrespective of
orbital periapse, as long as $r_{p}\!<\! q$. Thus, the star is destroyed
in a short time, e.g. $P_{0}(r_{h})\!\sim\!10^{5}$ yr in the Galactic
Center (GC).

In contrast, the other processes listed above proceed gradually over
an inspiral time $t_{0}\gg P_{0}$, as a small fraction of the orbital
energy is dissipated every peri-passage. The inspiral time typically
rises steeply with the periapse. In order for the extracted orbital
energy to heat the disk, tidally heat the star or power a high luminosity
of gravity waves, the star has first to decay into a short period
orbit.

Novikov et al. (\cite{Nov92}) estimate that tidal capture by a MBH
occurs for orbits with $r_{p}/r_{t}\!<\! b_{0}\!\sim\!3$. It then
follows that stars are scattered into tidal capture orbits at a rate
$b_{0}\!-\!1\!\sim\!2$ times faster than that for prompt tidal disruption
orbits. The orbital energy the star has to lose to circularize far
exceeds its own binding energy, so it is likely that it will ultimately
be disrupted (Rees \cite{Ree88}; AM03). This has led several authors
(Frank \& Rees \cite{Fra76}; Novikov et al. \cite{Nov92}; Magorrian
\& Tremaine \cite{Mag99}) to suggest that slow tidal inspiral may
be at least as important as prompt disruption for feeding the MBH
and for producing tidal flares. Simulations (Murphy et al. \cite{Mur91};
Freitag \& Benz \cite{Fre02a}) indicate that prompt tidal disruptions
supply between $\sim\!0.15$ to $0.65$ of the total mass of a low
mass MBH ($m\!\lesssim\!10^{7}\,\Mo$) in a low-density galactic nuclear
core. If the contribution of inspiraling stars were indeed as high
as implied by the ratio of the cross-sections, this would have far-reaching
implications: stars could supply most or even all of the MBH mass,
thereby establishing a direct link between $m$ and stellar dynamics
on a scale of $r_{h}$. 

However, a small initial periapse does not in itself guarantee ultimate
disruption. The star must also have enough time to complete its orbital
decay. In this Letter we revisit the questions: what is the time available
for orbital decay, and what is the inspiral event rate?

\section{Calculations}

\subsection{Scattering into inspiraling orbits}

We follow the analysis of Miralda-Escud\'{e} \& Gould (\cite{Mir00},
MG00) of the infall of a single mass population of stellar BHs into
a MBH by diffusion into the loss-cone due to 2-body encounters in
a Keplerian potential. 

Stars scattered from a volume $\mathrm{d}V$ to a periapse $r_{p}$
will spiral into the MBH in a time $t_{0}(a,r_{p})$, determined by
the dissipation process. The scattering rate per volume of stars into
orbits with periapse $<\! r_{p}$, $\mathrm{d}\Gamma(<\! r_{p})/\mathrm{d}V$,
can be estimated for the steady state distribution function (DF) (MG00
Eq.~28). The total rate, $\Gamma(<\! r_{p})\!=\!\int\mathrm{d}V\,\mathrm{d}\Gamma(<\! r_{p})/\mathrm{d}V$,
is obtained by integrating over the volume between an inner boundary
$r_{0}$ where the stellar cusp is truncated (e.g by stellar collisions)
to an outer boundary $r_{c}$. The outer boundary is set by requiring
that $t_{0}\!\leq\! t_{p}$, where $t_{p}(a,r_{p})\!\equiv\! r_{p}/\langle\mathrm{d}r_{p}/\mathrm{d}t\rangle$
is the orbit-averaged time for a change of order unity in the periapse
by diffusion in velocity space due to many small angle deflections. 

Typically, $t_{0}$ rises with $a$ while $t_{p}$ falls (Eq.~\ref{eq:tr}),
and so there is a critical semi-major axis $a_{c}(r_{p})$ such that
$\left.t_{0}(a_{c},r_{p})\right/t_{p}(a_{c},r_{p})=1$. Stars originating
from an orbit with $a_{0}\!>\! a_{c}$ do not have time, \emph{on
average}, to complete the inspiral. This simple condition is actually
too restrictive because $a(t)$ shrinks with time. Since stars are
scattered off the orbit at a rate of $\sim\! t_{p}^{-1}$, the Poisson
probability for avoiding this is $w\!=\!\exp(-s)$, where \begin{equation}
s(a_{c},r_{p})\equiv\int_{0}^{t_{0}(a_{c},r_{p})}\mathrm{d}t/t_{p}\left[a(t),r_{p}\right]\,.\label{eq:s}\end{equation}
 The critical semi-major axis is obtained by solving $s\!=\!1$ for
$a_{c}(r_{p})$. Formally, $\Gamma(<\! r_{p})$ should be evaluated
by the weighted integral $\int\mathrm{d}V\, w\,\mathrm{d}\Gamma(<\! r_{p})/\mathrm{d}V$
over all space. However $w$ falls off exponentially, and so $\Gamma(<\! r_{p})$
is well approximated by taking $w\!=\!1$ and $r_{c}(r_{p})\!=\!3a_{c}/2$,
the time-averaged radius on an $e\!\rightarrow\!1$ Keplerian orbit.

We assume here a simple power-law stellar DF, $n_{\star}\!\propto\! r^{-\alpha}$.
In the limit $e\rightarrow1$ (MG00 Eqs.~15--18, 21)

\begin{equation}
t_{p}(a)=A_{\alpha}\left(\frac{m}{\Ms}\right)^{2}\frac{P(a)}{N_{h}\log\Lambda_{1}}\left(\frac{r_{p}}{a}\right)\left(\frac{r_{h}}{a}\right)^{3-\alpha}\,,\label{eq:tr}\end{equation}
 where $N_{h}$ is the number of stars within $r_{h}$, $\Lambda_{1}\!=\!\Lambda(r_{p}/r_{c})^{1/4}$,
$\Lambda\!=\! m/M$ and\begin{equation}
A_{\alpha}\equiv\frac{15}{2^{5-\alpha}}\frac{(\alpha-1/2)!(9/2-\alpha)!}{(3-\alpha)(22-5\alpha)\alpha!(3-\alpha)!}\,.\label{eq:Wt}\end{equation}
 Note that $r_{p}/a\!=\!1-e\!=\!\vartheta^{2}$, where $\vartheta$
is the opening angle of the loss cone at $r\!=\! a$, so Eq.~(\ref{eq:tr})
is similar to the estimate $t_{\mathrm{scatter}}(a)\vartheta^{2}$
of Sigurdsson \& Rees (\cite{Sig97a}).

The volume contributing to $\Gamma(<\! r_{p})$ includes points at
distance $r$ far from the MBH, where $t_{0}\!>\! t_{p}$ for an orbit
with periapse $r_{p}$, but where $t_{0}\!\leq\! t_{p}$ is possible
for $r_{p}'<r_{p}$ due to higher dissipation at smaller periapse.
We generalize $\Gamma(<\! r_{p})$ to account for this by solving
$s(r,r'_{p})\!=\!1$ (Eq.~\ref{eq:s}) for $r'_{p}(r)$, and setting
$r_{m}(r)=\min[r_{p},r'_{p}(r)]$. The scattering rate is then (Eqs.
MG00 15--18, 28), \begin{equation}
\!\Gamma_{m}(<\! r_{p})\!=\! B_{\alpha}\frac{\Ms^{2}}{m^{2}}\frac{N_{h}^{2}}{P_{h}}\int_{r_{0}}^{r_{q}}\!\left[\frac{\log\Lambda}{\log(r/r_{m})}\!-\!\frac{1}{4}\right]\!\left(\frac{r}{r_{h}}\right)^{\gamma}\!\frac{\mathrm{d}r}{r_{h}}\,,\label{eq:Grp}\end{equation}
 where $\gamma\!\equiv\!7/2-2\alpha$, $P_{h}\!\equiv\! P(r_{h})$,
$r_{q}\!=\! r_{c}(q)$ is the maximal distance for prompt infall,
and \begin{equation}
B_{\alpha}\equiv\frac{4}{15}\sqrt{\frac{\pi}{2}}\frac{(3-\alpha)^{2}(10\alpha-1)\alpha!}{(\alpha-1/2)!}\quad(\alpha>\frac{1}{2})\,.\label{eq:WQ}\end{equation}
 Thus, the rate of \emph{successful} inspiral events is \begin{equation}
\Gamma_{i}(<\! r_{p})=\Gamma_{m}(<\! r_{p})-\Gamma_{m}(<\! q)\,.\end{equation}

The rate for prompt infall, $\Gamma_{p}(<\! r_{p})$, is that at which
stars are deflected into orbits with periapse $<\! r_{p}$ and reach
$r\!<\! r_{p}$ at least once, but do not necessarily finish the inspiral.
$\Gamma_{p}(<\! r_{p})$ thus includes horizon crossing, tidal disruption,
inspiral and tidal scattering. The prompt infall time $t_{0}\!\sim\! P_{0}$
does not depend on $r_{p}$ and so $r_{m}\!=\! r_{p}$ at all $r$
(for $\alpha\!<\!3$; Eqs.~\ref{eq:tr}, \ref{eq:acinfall}). The
rate $\Gamma_{p}(<\! r_{p})\!=\!\Gamma_{m}(<\! r_{p})$ is then \begin{eqnarray}
\Gamma_{p}(<\! r_{p}) & = & B_{\alpha}\left(\Ms/m\right)^{2}\left(N_{h}^{2}/P_{h}\right)\!\times\\
 &  & \!\!\!\!\!\!\!\!\!\!\!\!\!\!\!\!\!\!\!\!\!\!\!\!\!\left.\left\{ \log\Lambda\left(r_{p}/r_{h}\right)^{1+\gamma}\!\textrm{Ei}\!\left[\left(1\!+\!\gamma\right)\log\!\frac{r}{r_{p}}\right]\!-\!\frac{(r/r_{h})^{1+\gamma}}{4(1+\gamma)}\right\} \right|_{r_{0}}^{r_{c}}\,,\nonumber \end{eqnarray}
where $\mathrm{Ei}(x)\!\equiv\!-\int_{-x}^{\infty}\mathrm{d}te^{-t}/t$
is the exponential integral.

\subsection{Inspiral and infall timescales}

\emph{Tidal heating}.---We apply these results to {}``Hot Squeezars''
(HS, AM03), tidally heated stars that dissipate the heat on the surface.
We denote by a tilde dimensionless quantities in units of $G$=$\Ms$=$\Rs$=1
($\Ms$, $\Rs$ are the initial stellar mass and radius). In these
units $\Urt\!\simeq\!\UR(\Ut)\Um^{1/3}$. The HS inspiral time in
terms of $b\!\equiv\!\Urp/\Urt(0)$ is \begin{equation}
\Ut_{0}=(2\pi)^{2/3}\Um^{2/3}b^{6}\UP_{0}^{1/3}/T(b^{3/2})\qquad(\Um\gg1),\label{eq:t0sq}\end{equation}
where $T$ is the tidal coupling coefficient. Two models are used
to estimate the tidal heating: normal mode expansion (Press \& Teukolsky
\cite{Pre77}) and the affine stellar model (Carter \& Luminet \cite{Car85}).
For the former, $T$ is the leading multipole term, which is evaluated
numerically for a solar model (Alexander \& Kumar \cite{Ale01a}).
For the latter, we use the analytic approximation of Novikov et al.
(\cite{Nov92}). The tidal energy deposited in the star per peri-passage
is $\UdE\!=\! T/b^{6}\!\equiv\!\mathrm{const}$ (AM03). The orbital
evolution is (Eqs.~\ref{eq:s}--\ref{eq:tr}, AM03 Eq. 4) \begin{equation}
\Ua(\Ut)=\Ua_{0}\left(1-\Ut/\Ut_{0}\right)^{2}\,,\label{eq:at}\end{equation}

\begin{equation}
\Ut_{p}[\Ua(\Ut)]=\Ut_{p}(\Ua_{0})(1-\Ut/\Ut_{0})^{2\alpha-5}\,,\label{eq:tr0}\end{equation}

\begin{equation}
s=\left.\Ut_{0}\right/\left[2(3-\alpha)\Ut_{p}(\Ua_{0})\right]\,.\label{eq:q0}\end{equation}
 The critical semi-major axis is 

\begin{equation}
\Ua_{c}=\left[\frac{2(3-\alpha)A_{\alpha}T(b^{3/2})\Um^{4/3}}{\log\Lambda_{1}b^{5}N_{h}}\right]^{1/(3-\alpha)}\Urh\,.\label{eq:t0GW}\end{equation}

{}``Cold Squeezars'' (CS, AM03) dissipate the tidal energy in their
bulk and expand at constant effective temperature. The CS orbital
evolution is calculated below numerically. 

\emph{Gravitational waves}.---The time for inspiral by GW emission
($b\!>\! q/r_{t}$) is (Peters \cite{Pet64})

\begin{equation}
\Ut_{0}\simeq\frac{24\sqrt{2}\widetilde{c}^{5}}{85(2\pi)^{1/3}}\frac{b^{7/2}\UP_{0}^{1/3}}{\Um^{2/3}}\qquad(e\!\rightarrow\!1,\,\Um\!\gg\!1)\,,\end{equation}
where $\widetilde{c}$ is the speed of light. Inspiraling GW emitters,
like HSs, follow $\mathrm{d}a/\mathrm{d}t\!\propto\!\sqrt{a}$, so
Eqs. (\ref{eq:at})--(\ref{eq:q0}) apply and

\begin{equation}
\Ua_{c}=\left[\frac{85\pi(3-\alpha)A_{\alpha}\Um^{8/3}}{6\sqrt{2}\Uc^{5}\log\Lambda_{1}b^{5/2}N_{h}}\right]^{1/(3-\alpha)}\Urh\,.\end{equation}

\emph{Prompt infall}.---The mean infall time is $\Ut_{0}\!=\!\UP_{0}/4$,
and\begin{equation}
\Ua_{c}=\left[\frac{4A_{\alpha}\Um^{7/3}}{\log\Lambda_{1}N_{h}}\frac{b}{\Urh}\right]^{1/(4-\alpha)}\Urh\,.\label{eq:acinfall}\end{equation}

\section{Results}

We apply these results to the GC by modeling it as a power-law DF
with $m\!=\!2.6\!\times\!10^{6}\,\Mo$, $r_{h}\!=\!1.8\,\mathrm{pc}$
and $\alpha\!=\!1.8$, based on the empirically derived mass model
of Sch\"odel et al. (\cite{Sch02}). We assume a single mass population
and $N_{h}\!=\!\Um$. We consider 3 simple cases. (1) $1\,\Mo$ stars,
a high mean mass motivated by theoretical arguments and observational
evidence for a {}``top heavy'' initial mass function in the GC (e.g.
Morris \cite{Mor93}; Figer et al. \cite{Fig99}). This model is used
to test tidal inspiral and tidal scattering. (2) $0.1\,\Mo$ main-sequence
(MS) stars, which are the most resilient against tidal disruption
(Freitag \& Benz \cite{Fre02a}). This model is used to estimate GW
inspiral from MS stars. (3) $0.6\,\Mo$ stars. Of these, 10\% are
white dwarfs (WD) with $\Rs\!=\!0.01\,\Ro$, as is expected in old,
bulge-like stellar populations. The rest are MS stars, whose tidal
disruption radius is too large for efficient GW emission. This model
is used to estimate the rate of GW inspiral by WDs, for comparison
with previous works.

\subsection{Survival probability of tidally scattered stars}

Tidal inspiral is complementary to tidal scattering (Alexander \&
Livio \cite{Ale01b}), where stars narrowly escape tidal disruption
by being scattered to wider orbits, after suffering extreme tidal
distortion, spin-up, mixing and mass-loss that may affect their evolution
and appearance. Such stars eventually comprise a few percent of the
population within $r_{h}$. We now show that their probability of
survival is $\sim\!1$ by comparing the inspiral and prompt infall
rates. Table (\ref{tbl:inspiral}) lists the tidal inspiral rate $\Gamma_{i}(<\! b_{0}r_{t})$
($b_{0}$ is the maximal possible periapse determined by $r_{c}(b_{0}r_{t})\!=\! r_{0}$)
for $1\,\Mo$ HSs and CSs with the tidal heating estimated using either
normal mode expansion or the affine model (accounting numerically
for the fact that $\Ut(\UR\!=\! b)\!<\!\Ut_{0}$ for small $b$).
The stellar cusp is truncated at $r_{0}\!\sim\!0.02$ pc, the radius
where stellar collisions destroy MS stars (Alexander \cite{Ale99}).
The prompt disruption rate for this GC model is $\Gamma_{p}(\!<\! r_{t})\!\sim\!9\!\times\!10^{-5}\,\mathrm{yr}^{-1}$,
consistent with previous estimates, $\Gamma_{p}(\!<\! r_{t})\!=\!5\!\times\!10^{-5}\,\mathrm{yr}^{-1}$
(Syer \& Ulmer \cite{Sye99}) and $\Gamma_{p}(\!<\! r_{t})\!\sim\!\mathrm{few}\!\times\!10^{-5}\,\mathrm{yr}^{-1}$
(Alexander \cite{Ale99}). \emph{}However\emph{,} the squeezar inspiral
(tidal capture) event rate \emph{is only $\sim\!0.05$ of the prompt
disruption rate}, \emph{and not $\lesssim\!2$ times larger,} as naively
implied by the ratio of the cross-sections\emph{.} 

The probability of successful inspiral for stars with periapse between
$r_{t}$ and $r_{p}$ is\begin{equation}
P_{i}(\!<\! r_{p})=\frac{\!\Gamma_{i}(\!<\! r_{p})}{\Gamma_{p}(\!<\! r_{p})-\Gamma_{p}(\!<\! r_{t})}\,,\end{equation}
 while the tidal scattering survival probability is $P_{s}\!=\!1\!-\! P_{i}$.
Our squeezar models have $P_{s}(\lesssim\! r_{t})\!\sim\!0.8$ to
$P_{s}(<\! b_{0}r_{t})\!\sim\!0.9$. We thus confirm that $P_{s}\!\sim\!1$,
as was anticipated from general arguments (Alexander \& Livio \cite{Ale01b}).
The MBH's Brownian motion, neglected here, may further increase $P_{s}$
for loosely bound ($b\!\sim\! b_{0}$) tidally scattered stars. 

A tidal scattering event is deemed {}``strong'' if $r_{p}$ is within
some adopted limit. Since $P_{s}\!\sim\!1$, the tidal scattering
and prompt disruption rates are related. The diffusive cross-section,
modeled here, roughly scales as $\sim\! r_{p}^{\delta}$, where $\delta\!\sim\!(9\!-\!4\alpha)/(8\!-\!2\alpha)\!=\!0.4$
for $\alpha\!=\!1.8$ (Eqs.~\ref{eq:Grp}, \ref{eq:acinfall}). Unbound
stars with isotropic velocities have $\delta\!=\!1$ (Hills \cite{Hil75}).
Since most tidally scattered stars originate between the diffusive
and the full loss cone (isotropic) regimes, where $E\!\sim\!0$ (Lightman
\& Shapiro \cite{Lig77}), realistically $\delta\!\sim\!0.4$--$1$.

\begin{minipage}[t]{0.95\columnwidth}

\twocoltablecap{tbl:inspiral}{Inspiral in the Galactic Center}


\begin{tabular}{p{2in}ccp{0.12in}p{0.2in}cc}
\multicolumn{1}{p{0.45in}}{ {\footnotesize Process$^{a}$}}&
{\footnotesize $\left.r_{0}\right.^{b}$}&
{\footnotesize $\left.\Gamma_{i}\right.$}&
{\footnotesize $\bar{n}$}&
{\footnotesize $b_{0}$}&
\multicolumn{1}{p{0.5in}}{{\footnotesize $\left.\overline{L}_{1}\right.$}}&
{\footnotesize $\left.\overline{P}_{1}\right.$}\tabularnewline
\multicolumn{1}{p{0.45in}}{}&
{\footnotesize pc}&
{\footnotesize yr$^{-1}$}&
&
&
\multicolumn{1}{p{0.2in}}{}&
{\footnotesize yr}\tabularnewline
\hline
\multicolumn{1}{p{0.45in}}{{\footnotesize HS (S)}}&
 {\footnotesize $2(-\!2)$}&
 {\footnotesize $3(-\!6)$}&
 {\footnotesize $0.2$}&
 {\footnotesize $2.1$}&
\multicolumn{1}{p{0.2in}}{{\footnotesize $170\,\Lo$}}&
{\footnotesize $4(3)$}\tabularnewline
\multicolumn{1}{p{0.45in}}{{\footnotesize HS (A)}}&
 {\footnotesize $2(-\!2)$}&
 {\footnotesize $5(-\!6)$}&
 {\footnotesize $0.4$}&
 {\footnotesize $3.6$}&
\multicolumn{1}{p{0.2in}}{{\footnotesize $150\,\Lo$}}&
{\footnotesize $6(3)$}\tabularnewline
\multicolumn{1}{p{0.45in}}{{\footnotesize CS (S)}}&
 {\footnotesize $2(-\!2)$}&
 {\footnotesize $4(-\!6)$}&
 {\footnotesize $0.2$}&
 {\footnotesize $2.8$}&
\multicolumn{1}{p{0.2in}}{{\footnotesize $200\,\Lo$}}&
{\footnotesize $5(3)$}\tabularnewline
\multicolumn{1}{p{0.45in}}{{\footnotesize CS (A)}}&
 {\footnotesize $2(-\!2)$}&
 {\footnotesize $7(-\!6)$}&
 {\footnotesize $0.4$}&
 {\footnotesize $4.4$}&
\multicolumn{1}{p{0.2in}}{{\footnotesize $170\,\Lo$}}&
{\footnotesize $7(3)$}\tabularnewline
\hline
\multicolumn{1}{p{0.45in}}{{\footnotesize MS GW}}&
 {\footnotesize $6(-\!3)$}&
 {\footnotesize $2(-\!7)$}&
{\footnotesize $0.2$}&
 {\footnotesize $7\,\widetilde{r}_{S}$}&
\multicolumn{1}{p{0.2in}}{{\footnotesize $2(35)\,\mathrm{erg/}s$}}&
{\footnotesize $1(1)$}\tabularnewline
\multicolumn{1}{p{0.45in}}{{\footnotesize WD GW}}&
 {\footnotesize $4(-\!4)$}&
 {\footnotesize $2(-\!7)$}&
 {\footnotesize $0.04$}&
 {\footnotesize $25\,\widetilde{r}_{S}$}&
\multicolumn{1}{p{0.3in}}{{\footnotesize $1(36)\mathrm{\, erg}/s$}}&
{\footnotesize $1(2)$}\tabularnewline
\hline
\multicolumn{7}{l}{{\footnotesize $^{a}$S: Solar normal mode expansion. A: Affine stellar
model. }}\tabularnewline
\multicolumn{7}{l}{{\footnotesize $^{b}$The collisional destruction radius at $t_{H}\!=\!10\,\mathrm{Gyr}$,
defined }}\tabularnewline
\multicolumn{7}{l}{{\footnotesize ~~here by $e^{-t_{H}/10t_{\mathrm{coll}}}\!=\!0.1$
(Murphy et al. \cite{Mur91}, Fig. 4a) }}\tabularnewline
\end{tabular}

\vspace{1ex}


\end{minipage}

\subsection{Inspiral in the Galactic Center}

The observable implications of having on average $\overline{n}$ inspiraling
stars near the MBH can be estimated by considering the properties
of the leading (shortest period) star. The mean number is $\overline{n}(<b_{0}r_{t})\!\equiv\!\int_{r_{t}}^{b_{0}r_{t}}\mathrm{d}r_{p}\left\langle t_{0}\right\rangle (\mathrm{d}\Gamma_{i}/\mathrm{d}r_{p})$,
where $\left\langle t_{0}\right\rangle \!\sim\!(2\!+\!2\gamma)t_{0}(r_{p},r_{c})/(3\!+\!2\gamma)$
is the $r$-averaged value of $t_{0}$ ($t_{0}\!\propto\!\sqrt{a_{0}}$
and Eq. \ref{eq:Grp}). The mean inspiral time $\overline{t}_{0}\!\equiv\!\overline{n}(<b_{0}r_{t})/\widetilde{\Gamma}_{i}(<b_{0}r_{t})$
and the corresponding averaged $\sqrt{\Ua_{0}}$ also define $\overline{a}_{0}$
and $\overline{P}_{0}$ for {}``typical'' HSs or GW emitters. For
simplicity, we adopt this estimate also for CSs, although their $t_{0}(a_{0})$
is different. The leading star completes on average $\overline{t}_{1}/\overline{t}_{0}\!=\!\overline{n}/\left(\overline{n}+1\right)$
of its inspiral time in $N\!=\!\overline{n}(\overline{n}\!+\!2)\overline{t}_{0}/2\overline{P}_{0}$
orbits (AM03). Since $\overline{a}_{1}\!=\!\overline{a}_{0}/(\overline{n}+1)^{2}$
and $\overline{P}_{1}\!=\!\overline{P}_{0}/(\overline{n}+1)^{3}$,
the total extracted orbital energy is $\Delta\overline{E}_{1}=(\Um/2\overline{a}_{0})\overline{n}\left(\overline{n}+2\right)$
and the tidal luminosity is $\overline{L}_{1}\left(\overline{t}_{1}\right)\sim\Delta\overline{E}_{1}/N\overline{P}_{1}=(\Um/\Ua_{0}\overline{t}_{0})(\overline{n}+1)^{3}$.
Typically, $\overline{L}_{1}\!\gg\Ls$ (the intrinsic stellar luminosity)
even for small $\overline{n}$.

Table (\ref{tbl:inspiral}) lists $\overline{n}$,  $\overline{L}_{1}$
and $\overline{P}_{1}$ for tidal and GW inspiral. We find that the
GC contains on average $\sim\!0.1$--1, squeezars, and that on average
the leading squeezar is $\sim\!5.5^{\mathrm{m}}$ brighter than its
normal bolometric magnitude. The rate of WD inspiral derived here,
$\Gamma_{i}\!\sim\!2\!\times\!10^{-7}\,\mathrm{yr}^{-1}$, agrees
with the estimates of Sigurdsson \& Rees (\cite{Sig97a}) and Freitag
(\cite{Fre03}), but the rate of GW inspiral by MS stars, $\Gamma_{i}\!\sim\!2\!\times\!10^{-7}\,\mathrm{yr}^{-1}$,
is less than $0.1$ of that estimated by Freitag (\cite{Fre03}).
The source of this discrepancy is unclear.

\section{Discussion and summary}

Different dissipation mechanisms may lead to the same outcome: orbital
decay around a MBH in the presence of two-body perturbations. In this
Letter we derive general expressions for estimating the inspiral event
rate for any given inspiral time $t_{0}(a_{0},r_{p})$ and orbital
evolution $a(t)$. Inspiral is a race between orbital energy extraction
and two-body scattering. The probability that a star will reach the
final, observationally interesting stage of a short period orbit is
small unless it starts out on a tight enough orbit. Since there are
fewer stars close to the MBH, inspiral events are much rarer than
prompt disruption events. 

We applied these results to the GC, using simple single-mass models
to represent the stellar cluster. These reproduce the prompt disruption
and the WD inspiral event rates that were independently estimated
by previous studies. We find that (1) The survival probability of
tidally scattered stars is $0.8$--$0.9$. (2) The rate of tidal scattering
scales with the prompt tidal disruption rate as $\Gamma_{p}(<\! r_{t})[(r_{p}/r_{t})^{\delta}-1]$
with $\delta\!\sim\!0.4$--1. (3) The contribution of slow tidal inspiral
in the GC to the total tidal disruption rate is only $\sim\!5\%$,
and not $100$--$200\%$ as proposed by previous studies. (4) The
GC contains on average $\sim\!0.1$--$1$ squeezars at any given time,
with a tidal bolometric luminosity $\gtrsim\!150\Ls$ on $\sim\!5\times\!10^{3}$
yr orbits. (5) There are on average $\sim\!0.1$ GW emitters in the
GC at any given time. 

The uncertainties in these estimates can be addressed by more detailed
modeling of the stellar cluster, which should include a realistic,
multi-mass stellar DF and take into account mass segregation. Other
dynamical processes not considered here, such as resonant scattering
(Rauch \& Tremaine \cite{Rau96}), deviations from spherical symmetry
(Magorrian \& Tremaine \cite{Mag99}), chaotic orbits in triaxial
systems (Poon \& Merritt \cite{Poo02}), or the effects of massive
perturbers (Zhao et al. \cite{Zha02}) may increase the inspiral event
rates. The MBH's Brownian motion will not have a large effect, as
typical inspiral orbits originate well within $r_{h}$, where the
stars follow the MBH (Reid et al. \cite{Rei03}).

\acknowledgements{We thank M. Freitag and M. Morris for helpful discussions. TA is
supported by ISF grant 295/02-1, Minerva grant 8484, and a grant by
Sir H. Djangoly, CBE, of London, UK.}

\end{document}